\documentclass[aps,prl,reprint,groupedaddress]{revtex4-2}
\usepackage{amssymb}
\usepackage{bm}
\usepackage{verbatim}
\usepackage{graphicx}
\usepackage[usenames,dvipsnames]{color} 
\usepackage{hyperref}

\begin{document}


\title{Topological Entropy of Two Dimensional Turbulence}


\author{Amal Manoharan, Sai Subramanian and Ashwin Joy}
\email[]{ashwin@physics.iitm.ac.in}
\affiliation{Department of Physics, Indian Institute of Technology Madras, Chennai - 600036}


\date{\today}

\begin{abstract}
  Deformation of material lines drives transport and dissipation in many industrial and natural flows. Here we report an exact Eulerian formula for the stretching rate of a material line, also known as the topological entropy,  in a prototype two-dimensional fluid. The only requirement is a distribution of eigenvalues of the strain rate tensor and their decorrelation time. This eliminates the need for Lagrangian tracking in experimental turbulence where particle trajectories are entangled, and thus poorly resolved. Numerical simulations reveal an  excellent agreement between our  Eulerian estimate and the stretching rate of a Lagrangian material line, over a wide range  of Reynolds number.

\end{abstract}


\maketitle
About a century ago, G.I. Taylor performed a remarkably simple yet thought provoking experiment on a viscous fluid confined in the region between two rotating cylinders \cite{taylor1923viii}. He showed that as the relative rotation rate becomes sufficiently high, the steady vortices in the fluid become irregular and chaotic. This emergent complexity in the fluid state \textemdash seemingly infinite \textemdash is often deemed to be the hallmark of turbulence. It seems natural that a quantitative understanding of this complex state will form the basis of any reliable prediction of material transport and dissipation in many natural and industrial flows that are inherently turbulent \cite{roberts1923theoretical,wang2012advanced,jicha2000dispersion,fernando2010flow,abuhegazy2020numerical,tel2005chemical}. Within dynamical systems, this is usually addressed through the topological entropy \cite{adler1965topological,ott2002chaos,newhouse1993estimation} \textemdash a non-negative extended  number that serves as a measure of complexity. In fluids, this is determined as the stretching rate of a material curve made up of tracer particles that are advected by the flow. Notable applications range from microfluidics \cite{doi:10.1098/rsta.2003.1360,lee2011microfluidic} through industrial \cite{finn2011topological, chu2018topological,thiffeault2018mathematics} to oceanic \cite{10.1063/1.3262494,candelaresi2017quantifying,klapper1995rigorous,olascoaga2012forecasting} and atmospheric \cite{Hazpra-S,haszpra2011volcanic} flows. Explicitly, one can write the topological entropy as 
\begin{equation}
  \mathcal{S} = \frac{1}{{\mathcal{T}}}\ln \bigg(\frac{l_{\mathcal{T}}}{l_0}\bigg)
  \label{S-rate}
\end{equation}
where $l_0$ is the original length of a material curve that deforms to a length $l_{\mathcal{T}}$ over some time $\mathcal{T}\gg 1/\mathcal{S}$ to ensure that the curve has visited a large number of unstable cycles. The idea seems straightforward to implement in numerical simulations; however, it presents a significant technical difficulty in laboratory experiments \textemdash one needs to continuously monitor a large number of particle trajectories that are progressively entangled by the turbulent background \cite{westerweel2013particle,10.1063/5.0211508}.  For example, modern Lagrangian measurement techniques are limited to the best combination of the large eddy length scale of $0.1$m and the Kolmogorov time scale of $10\mu$s. This puts only a small class of laboratory flows within the ambit of these techniques and leaves out many natural flows such as atmospheric  and oceanic systems, where these scales are at least three orders of magnitude larger \cite{toschi2009lagrangian}. Although this can be partly alleviated by considering a limited number of particle trajectories \cite{braid-2D}, it is in the general nature of the Lagrangian description where the problem remains formidable. Here we aim to eliminate the need for Lagrangian tracking by presenting an exact Eulerian framework that only requires a spatial distribution of the eigenvalues of the strain rate tensor and their decorrelation times \textemdash both easily obtained from wire sensors placed in the turbulent flow. The predicted entropy is in excellent agreement with the stretching rate of Lagrangian material lines observed in numerical simulations of inertial turbulence. In what follows, we give the  derivation of our exact formula in a two-dimensional setting that motivates an application to geophysical flows.
\begin{figure}[b]
  \centering
  \includegraphics[width=0.8\linewidth,keepaspectratio]{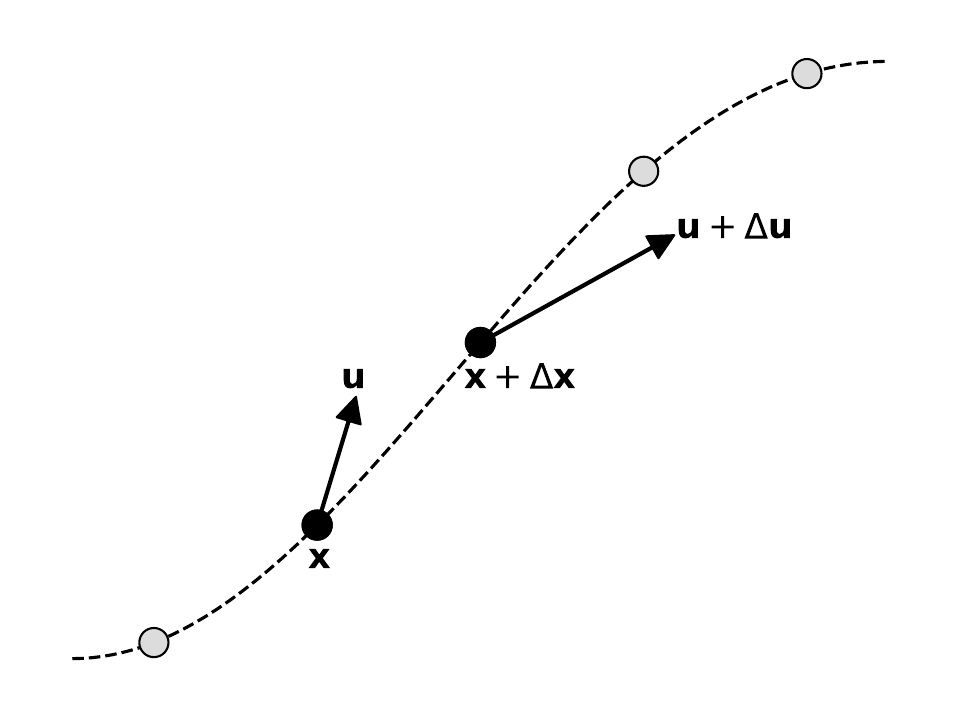} \caption{Schematic showing the relative motion of two nearby passive tracers that make up a Lagrangian material line. The distance between the pairs is exaggerated for clarity. In practice, they are close enough for Eq. (\ref{gradv2}) to hold.}
  \label{fig:pair}
\end{figure}

\begin{figure*}
  \includegraphics[width=0.95\linewidth,keepaspectratio]
  {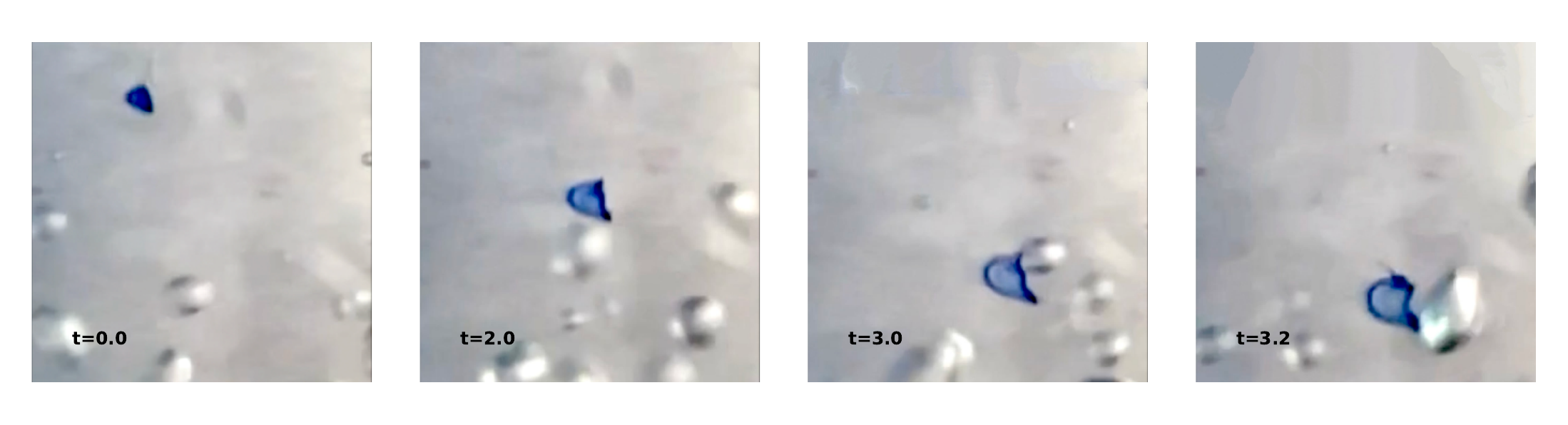}
  \includegraphics[width=0.95\linewidth,keepaspectratio]
  {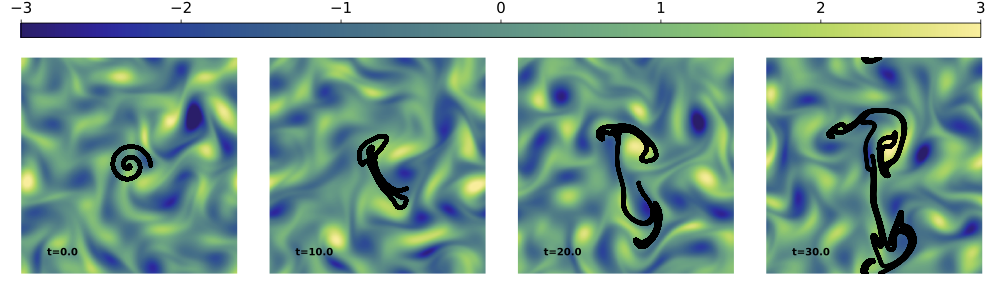}
  \caption{Top row: A trail of blue ink released in a stream of water flowing from left to right  is shown at different times (measured in seconds). The material loop is deformed and stretched by the background flow before it breaks apart further downstream. Bottom row: Time evolution of a material curve made up of passive tracers in a typical numerical simulation of two-dimensional inertial turbulence; color bar shows the local fluid vorticity and time is measured in dimensionless units. The exponential stretching rate of this curve is independent of its initial shape, and provides a Lagrangian estimate of the topological entropy that is plotted in Fig. \ref{inertial-S}.}
  \label{fig:tracers}
\end{figure*}

\textit{Theory:} Consider a pair of passive tracers separated by a vector $\Delta \bm{x}$ as shown in Fig. \ref{fig:pair}. Taking the velocity at one of them as $\bm u$, the velocity at the other can be approximated to within linear order as  $\bm u + \nabla \bm u \cdot \Delta \bm x $. Here $\nabla \bm u$ denotes the velocity gradient tensor that can be taken as the sum of the \textit{symmetric} strain rate tensor $\bm s $ and the \textit{antisymmetric} rotation tensor $\bm r$; written below
\begin{equation}
  \nabla \bm u = \bm{s} + \bm r = \frac{1}{2} \bigg(\frac{\partial u_i}{\partial x_j} + \frac{\partial u_j}{\partial x_i}\bigg) + \frac{1}{2} \bigg(\frac{\partial u_i}{\partial x_j}  - \frac{\partial u_j}{\partial x_i}  \bigg).
  \label{gradv}
\end{equation}
This gives the velocity difference between the pair as
\begin{equation}
  \Delta \bm u = \nabla \bm u \cdot \Delta \bm x = \bm{s} \cdot \Delta \bm x + \frac{1}{2} (\bm \omega \times  \Delta \bm x)
  \label{gradv2}
\end{equation}
where $\bm \omega = \bm \nabla \times \bm u$ is the local vorticity. Note that the contribution from local rotation is orthogonal to both the separation vector and local vorticity. A direct consequence of this is that local rotation cannot change the separation distance between the tracers, only strain can do that. Explicitly speaking, 
\begin{equation}
  \frac{\text{d}}{\text{d}t}|\Delta \bm x|^2 = 2 \Delta \bm x \cdot \Delta \bm u = 2 \Delta \bm x \cdot \bm{s} \cdot \Delta \bm x
  \label{dxdt}
\end{equation}
Therefore, in the eigenbasis of the strain rate tensor, the separation vector must evolve as 
\begin{equation}
\frac{\text{d}}{\text{d}t} \Delta \bm x = \bm{s} \cdot \Delta \bm x  
\label{dx-sol}
\end{equation}
with the solution in component form at some time $\tau$ as 
\begin{eqnarray}
  \Delta x &=& \Delta x_0 e^{\lambda \tau} \nonumber\\
  \Delta y &=& \Delta y_0 e^{-\lambda \tau}.
  \label{dxdy}
\end{eqnarray}
Here, $\lambda>0$ is an eigenvalue of the strain rate tensor. Note that expansion along one direction naturally leads to compression along the other due to incompressibility ($\text{tr}(\bm{s}) = 0$). If the separation vector is initially resolved as $\Delta x_0 = d_0 \cos \theta$ and $\Delta y_0 = d_0 \sin \theta$, then the ratio 
\begin{equation}
  \frac{\sqrt{\Delta x^2 + \Delta y^2}}{d_0} = \sqrt{e^{2 \lambda \tau} \cos^2\theta + e^{-2 \lambda \tau} \sin^2 \theta}  = f(\lambda, \theta)
  \label{scaling}
\end{equation}
defines the scaling function over the  time interval $\tau$. This idea can be exploited to track a material line composed of $n$ tracer pairs, each initially separated by $d_0$, as it deforms over an arbitrary time $\mathcal{T} = m\tau$ ($m \gg 1$). Put simply, the deformed length can be written as 
\begin{equation}
  l_{\mathcal{T}} = d_0 \sum_{i=1}^n \prod_{\alpha=1}^m f(\lambda_{i\alpha}, \theta_{i\alpha})
  \label{curve-length}
\end{equation}
The topological entropy is then  computed as 
\begin{eqnarray}
  \mathcal{S} = \frac{1}{m\tau} \ln \bigg[\frac{l_{\mathcal{T}}}{nd_0} \bigg] &=& \frac{1}{m\tau} \ln \bigg[\frac{1}{n}\sum_{i=1}^n \prod_{\alpha=1}^m f(\lambda_{i\alpha}, \theta_{i\alpha})\bigg]\nonumber\\
  &=& \frac{1}{m\tau} \ln \bigg\langle \prod_{\alpha=1}^m f(\lambda_{\alpha}, \theta_{\alpha})\bigg\rangle_{\text{pairs}}
  \label{LagrangianS}
\end{eqnarray}

\begin{figure}[]
  \centering
  \includegraphics[width=\linewidth,keepaspectratio]{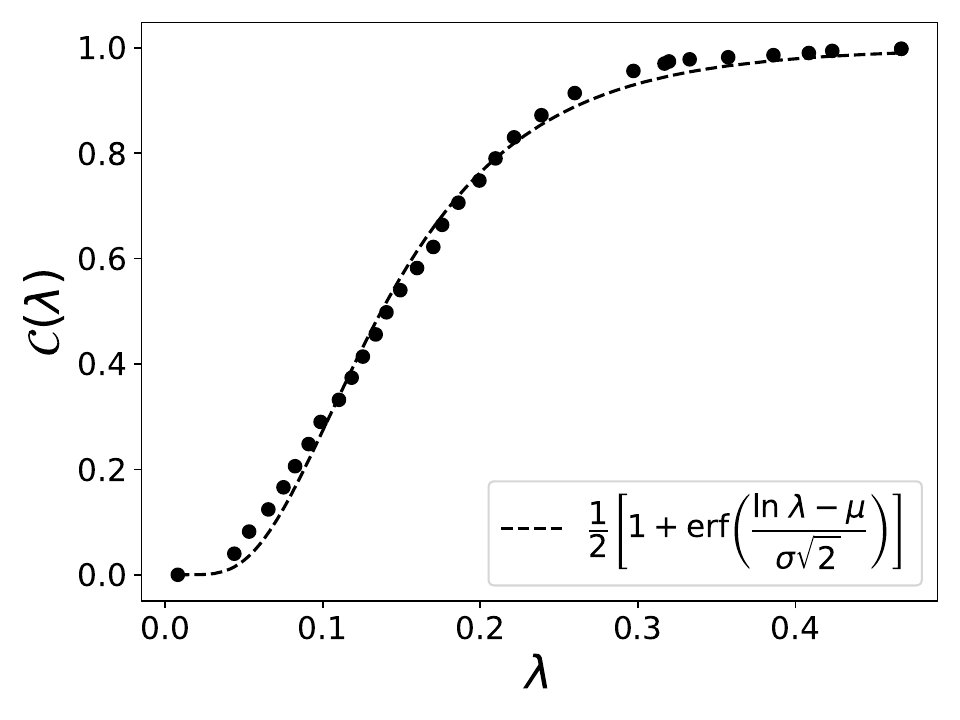}
  \includegraphics[width=\linewidth,keepaspectratio]{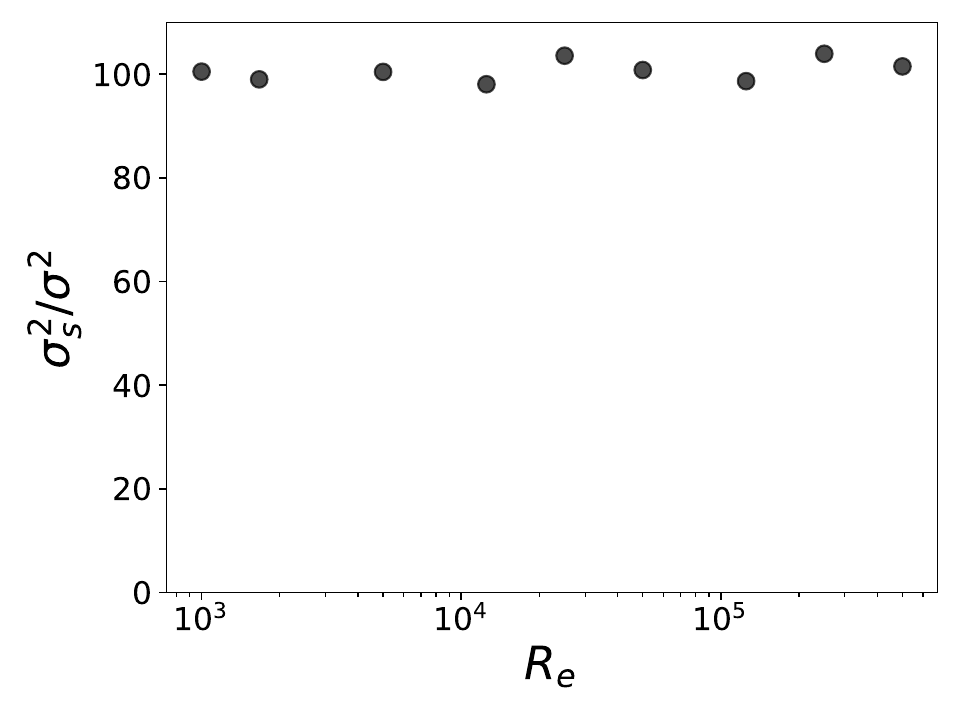}
  \caption{Top: Cumulative distribution function $\mathcal{C}(\lambda)$ of the eigenvalue of the strain rate tensor for a representative $R_{\text{e}} = 5000$. The fit suggests that the distribution is indeed log-normal with the associated mean $\mu = -2.00$ and standard deviation $\sigma = 0.56$. Bottom: The observed ratio of the two variances after a waiting time of $100 \tau$ verifies Eq. (\ref{var-strain}).}
  \label{fig:sigSq}
\end{figure}

We must emphasize here that the time-accumulated scaling function  $\prod_{\alpha=1}^m f(\lambda_{\alpha}, \theta_{\alpha})$ is log-normally distributed due to the positivity of each factor $f(\lambda_{\alpha}, \theta_{\alpha})$ and the central limit theorem. This allows us to move the average outside the logarithm and write 
\begin{eqnarray}
  \mathcal{S} &=&  \frac{1}{m\tau} \bigg\langle \ln \bigg(\prod_{\alpha=1}^m f(\lambda_{\alpha}, \theta_{\alpha})\bigg) \bigg\rangle_{\text{pairs}} + \frac{\sigma^2_s}{2m\tau} \nonumber\\
  &=& \frac{1}{\tau} \langle \ln ( f(\lambda, \theta )\rangle_{\text{pairs, time}} +  \frac{\sigma^2_s}{2m\tau} 
  \label{s-approx}
\end{eqnarray}
with $\sigma^2_s$ as the associated variance of $\prod_{\alpha=1}^m f(\lambda_{\alpha}, \theta_{\alpha})$. As the eigenvalue $\lambda$ is lognormally distributed with an associated variance $\sigma^2$ (see Fig. \ref{fig:sigSq}: Top), the individual scaling factor $f(\lambda, \theta) \approx 1 + \lambda \tau$ is also lognormally distributed with the same associated variance. Thus, the two associated variances must be related as (see Fig. 3: Bottom)
\begin{equation}
  \sigma^2_s \approx m \sigma^2
  \label{var-strain}
\end{equation}
This simplifies the entropy expression to 
\begin{equation}
  \mathcal{S} = \frac{1}{\tau} \langle \ln ( f(\lambda, \theta )\rangle_{\text{pairs, time}} + \frac{\sigma^2}{2\tau}
  \label{S-Lagrangian}
\end{equation}
Equation (\ref{S-Lagrangian}) gives a prescription to compute $\mathcal{S}$ by a Lagrangian average $\langle \rangle_{\text{pairs, time}}$ that is often difficult to obtain in turbulent flows. However, an important breakthrough is possible if we take $\tau$ as the decorrelation timescale of the eigenvalue $\lambda$. The constituent tracer pairs of the material line are then expected to sample through a distribution of $\lambda$ and $\theta$ over a waiting time of $\mathcal{T} \gg \tau$, effectively allowing us to rewrite the Lagrangian average 
\begin{equation}
  \langle \cdots \rangle_{\text{pairs}, \text{time}} \equiv \langle \cdots \rangle_{\lambda, \theta}
  \label{}
\end{equation}
where the averages over $\theta$ and $\lambda$ can be performed independently. Since $\theta$ is uniformly distributed between $0$ and $\pi/2$, the $\theta$-average of $\ln f(\lambda, \theta )$ can be worked out through a contour integration (see supplemental material \cite{SM}) as 
\begin{equation}
  \frac{2}{\pi}\int_0^{\pi/2} \ln f(\lambda, \theta)  \text{d}\theta = \ln (\cosh(\lambda \tau))
  \label{theta-avg}
\end{equation}
Using this in Eq. (\ref{S-Lagrangian}), we get an exact formula
\begin{equation}
  \mathcal{S} = \frac{1}{\tau}\langle\ln(e^{\sigma^2/2}\cosh(\tau\lambda)) \rangle
  \label{final-S}
\end{equation} 
for the topological entropy under the Eulerian framework. The only input to this formula is a distribution of the eigenvalue $\lambda$ at fixed locations in the fluid that can be  obtained from traditional Eulerian probes \textemdash the need for Lagrangian tracking is thus removed. The decorrelation time $\tau$ can be extracted from the e-folding of the auto-correlation function, 
\begin{equation}
  \mathcal{C}(t) = \langle (\lambda(t) - \overline{\lambda(t)}) (\lambda(0) - \overline{\lambda(0)})\rangle
  \label{auto-corr}
\end{equation}
where $\overline{\lambda(t)}$ indicates a spatial average of $\lambda$ at any time $t$. Armed with the Eulerian framework in Eq. (\ref{final-S}), we will now validate our theory by performing direct numerical simulations of two-dimensional turbulent flows maintained at steady state. This is discussed next.

\begin{figure}
  \includegraphics[width=1.05\linewidth,keepaspectratio]{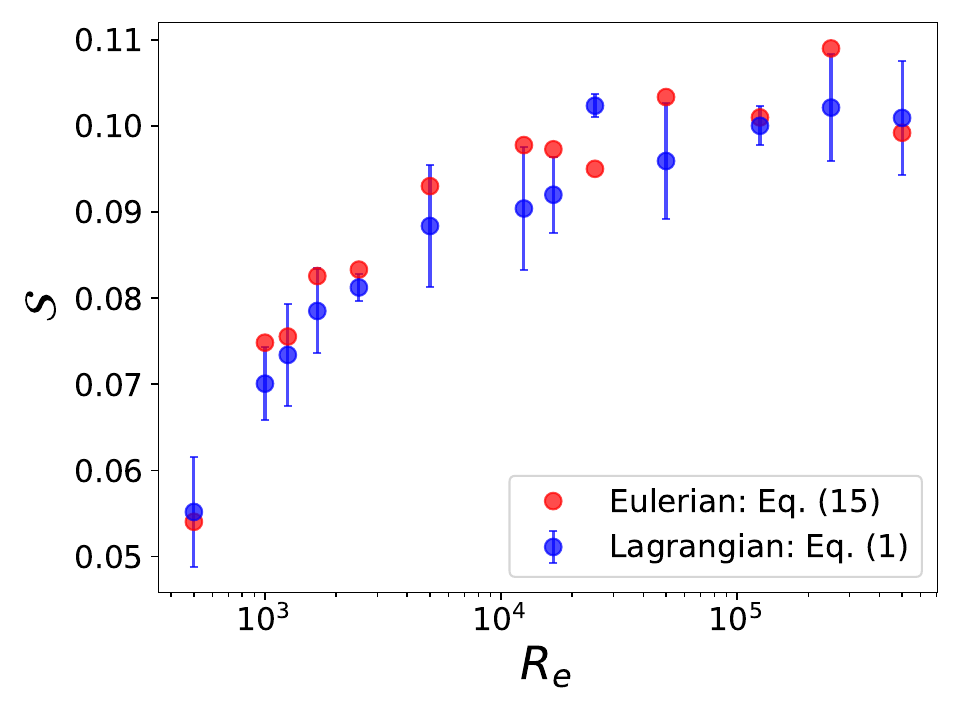}
  \caption{Eulerian and Lagrangian measures of topological entropy computed in simulations of 2D inertial turbulence. Error bars shown in the Lagrangian measure span over a unit standard deviation. The agreement between these two measures over three decades of $R_{\text{e}}$ demonstrates a proof of our calculation. Note the saturation of $\mathcal{S}$ at high $R_{\text{e}}$ is consistent with the energy spectrum progressively aligning with the Kraichnan scaling \cite{kraichnan1967inertial}.}
  \label{inertial-S}
\end{figure}
\textit{Numerical simulations:} We model a two-dimensional flow by the well known incompressible Navier-Stokes equation, which in the dimensionless form reads as,
\begin{eqnarray}
  \frac{\partial \omega}{\partial t} + \frac{\partial \Psi}{\partial y}\frac{\partial \omega}{\partial x} - \frac{\partial \Psi}{\partial x}\frac{\partial \omega}{\partial y} = \frac{1}{R_{\text{e}}} \bm\nabla^2 \omega - \alpha\omega + F
  \label{NS}
\end{eqnarray}
where the streamfunction $\Psi$ determines both vorticity $\omega = -\bm\nabla^2 \Psi$ and velocity $\bm u = \bm\nabla \times \Psi \bm{\hat{z}}$ of the fluid. The Reynolds number $R_{\text{e}}$ effectively characterizes the ratio of convective to viscous effects in the turbulent state. In addition to viscous dissipation at small scales, we include the Ekman friction $\alpha$ that acts as a sink of energy at large scales. To maintain a steady state, we balance the dissipation with a Markovian forcing $F$ that is localized over a narrow band of wave numbers $3\leq k \leq 7$ \cite{maltrud1991energy}. We take $\alpha = 0.01$ and solve Eq. (\ref{NS}) using a pseudospectral method over a doubly periodic square grid of $2048^2$ points with size $2\pi$. A time step of $0.001$ for integration gives compliance with the Courant-Friedrichs-Lewy condition at all values $R_e$.  At steady state, we initialize a material curve composed of passive tracers that are advected by the local flow over time, see Fig. \ref{fig:tracers}. The curve deforms over time and its exponential stretching rate provides a Lagrangian measure of the topological entropy that is shown in Fig. \ref{inertial-S} along with the Eulerian measure that was presented in Eq. \ref{final-S}. Impressively, the two measures are in excellent agreement over three decades of $R_{\text{e}}$, directly asserting our theoretical calculation. For an experimentalist, this holds a great prospect of minimizing the inputs to the entropy calculation to just the local velocity gradient tensor \textemdash a quantity that is easily obtained from traditional methods \cite{wallace2010measurement,tsinober1992experimental}. Thus we believe  our work provides great utility in experimental turbulence where the spatial resolution of particle tracking methods continues to be challenging. It would be interesting to extend the  core concept and results of this work to three-dimensional flows that will exhibit vortex stretching and yield an additional independent eigenvalue of the strain rate tensor. Our preliminary calculations suggest that the topological entropy will contain elliptic integrals and consequently, cannot be expressed in terms of regular functions. We defer a detailed analysis of this matter to a future communication as that is outside the scope of this paper. We close by remarking that there is another related tool to explore the dynamics of passive tracers, namely the Lyapunov exponent. However, unlike the topological entropy, its numerical determination requires particle trajectories to be infinitesimally close. Since a material line can a reach length of several thousands of kilometers in  geophysical flows, the topological entropy measured typically in days$^{-1}$ \cite{Hazpra-S}, clearly stands out as a better choice. 

\acknowledgements{We thank Sumesh Thampi for useful comments on the manuscript.}

\bibliography{manuscript}

\begin{thebibliography}{30}%
\makeatletter
\providecommand \@ifxundefined [1]{%
 \@ifx{#1\undefined}
}%
\providecommand \@ifnum [1]{%
 \ifnum #1\expandafter \@firstoftwo
 \else \expandafter \@secondoftwo
 \fi
}%
\providecommand \@ifx [1]{%
 \ifx #1\expandafter \@firstoftwo
 \else \expandafter \@secondoftwo
 \fi
}%
\providecommand \natexlab [1]{#1}%
\providecommand \enquote  [1]{``#1''}%
\providecommand \bibnamefont  [1]{#1}%
\providecommand \bibfnamefont [1]{#1}%
\providecommand \citenamefont [1]{#1}%
\providecommand \href@noop [0]{\@secondoftwo}%
\providecommand \href [0]{\begingroup \@sanitize@url \@href}%
\providecommand \@href[1]{\@@startlink{#1}\@@href}%
\providecommand \@@href[1]{\endgroup#1\@@endlink}%
\providecommand \@sanitize@url [0]{\catcode `\\12\catcode `\$12\catcode
  `\&12\catcode `\#12\catcode `\^12\catcode `\_12\catcode `\%12\relax}%
\providecommand \@@startlink[1]{}%
\providecommand \@@endlink[0]{}%
\providecommand \url  [0]{\begingroup\@sanitize@url \@url }%
\providecommand \@url [1]{\endgroup\@href {#1}{\urlprefix }}%
\providecommand \urlprefix  [0]{URL }%
\providecommand \Eprint [0]{\href }%
\providecommand \doibase [0]{https://doi.org/}%
\providecommand \selectlanguage [0]{\@gobble}%
\providecommand \bibinfo  [0]{\@secondoftwo}%
\providecommand \bibfield  [0]{\@secondoftwo}%
\providecommand \translation [1]{[#1]}%
\providecommand \BibitemOpen [0]{}%
\providecommand \bibitemStop [0]{}%
\providecommand \bibitemNoStop [0]{.\EOS\space}%
\providecommand \EOS [0]{\spacefactor3000\relax}%
\providecommand \BibitemShut  [1]{\csname bibitem#1\endcsname}%
\let\auto@bib@innerbib\@empty
\bibitem [{\citenamefont {Taylor}(1923)}]{taylor1923viii}%
  \BibitemOpen
  \bibfield  {author} {\bibinfo {author} {\bibfnamefont {G.~I.}\ \bibnamefont
  {Taylor}},\ }\bibfield  {title} {\bibinfo {title} {Viii. stability of a
  viscous liquid contained between two rotating cylinders},\ }\href@noop {}
  {\bibfield  {journal} {\bibinfo  {journal} {Philosophical Transactions of the
  Royal Society of London. Series A, Containing Papers of a Mathematical or
  Physical Character}\ }\textbf {\bibinfo {volume} {223}},\ \bibinfo {pages}
  {289} (\bibinfo {year} {1923})}\BibitemShut {NoStop}%
\bibitem [{\citenamefont {Roberts}(1923)}]{roberts1923theoretical}%
  \BibitemOpen
  \bibfield  {author} {\bibinfo {author} {\bibfnamefont {O.}~\bibnamefont
  {Roberts}},\ }\bibfield  {title} {\bibinfo {title} {The theoretical
  scattering of smoke in a turbulent atmosphere},\ }\href@noop {} {\bibfield
  {journal} {\bibinfo  {journal} {Proceedings of the Royal Society of London.
  Series A, Containing Papers of a Mathematical and Physical Character}\
  }\textbf {\bibinfo {volume} {104}},\ \bibinfo {pages} {640} (\bibinfo {year}
  {1923})}\BibitemShut {NoStop}%
\bibitem [{\citenamefont {Wang}\ \emph {et~al.}(2012)\citenamefont {Wang},
  \citenamefont {Lin},\ and\ \citenamefont {Chen}}]{wang2012advanced}%
  \BibitemOpen
  \bibfield  {author} {\bibinfo {author} {\bibfnamefont {M.}~\bibnamefont
  {Wang}}, \bibinfo {author} {\bibfnamefont {C.-H.}\ \bibnamefont {Lin}},\ and\
  \bibinfo {author} {\bibfnamefont {Q.}~\bibnamefont {Chen}},\ }\bibfield
  {title} {\bibinfo {title} {Advanced turbulence models for predicting particle
  transport in enclosed environments},\ }\href@noop {} {\bibfield  {journal}
  {\bibinfo  {journal} {Building and Environment}\ }\textbf {\bibinfo {volume}
  {47}},\ \bibinfo {pages} {40} (\bibinfo {year} {2012})}\BibitemShut {NoStop}%
\bibitem [{\citenamefont {Jicha}\ \emph {et~al.}(2000)\citenamefont {Jicha},
  \citenamefont {Pospisil},\ and\ \citenamefont
  {Katolicky}}]{jicha2000dispersion}%
  \BibitemOpen
  \bibfield  {author} {\bibinfo {author} {\bibfnamefont {M.}~\bibnamefont
  {Jicha}}, \bibinfo {author} {\bibfnamefont {J.}~\bibnamefont {Pospisil}},\
  and\ \bibinfo {author} {\bibfnamefont {J.}~\bibnamefont {Katolicky}},\
  }\bibfield  {title} {\bibinfo {title} {Dispersion of pollutants in street
  canyon under traffic induced flow and turbulence},\ }\href@noop {} {\bibfield
   {journal} {\bibinfo  {journal} {Environmental monitoring and assessment}\
  }\textbf {\bibinfo {volume} {65}},\ \bibinfo {pages} {343} (\bibinfo {year}
  {2000})}\BibitemShut {NoStop}%
\bibitem [{\citenamefont {Fernando}\ \emph {et~al.}(2010)\citenamefont
  {Fernando}, \citenamefont {Zajic}, \citenamefont {Di~Sabatino}, \citenamefont
  {Dimitrova}, \citenamefont {Hedquist},\ and\ \citenamefont
  {Dallman}}]{fernando2010flow}%
  \BibitemOpen
  \bibfield  {author} {\bibinfo {author} {\bibfnamefont {H.~J.}\ \bibnamefont
  {Fernando}}, \bibinfo {author} {\bibfnamefont {D.}~\bibnamefont {Zajic}},
  \bibinfo {author} {\bibfnamefont {S.}~\bibnamefont {Di~Sabatino}}, \bibinfo
  {author} {\bibfnamefont {R.}~\bibnamefont {Dimitrova}}, \bibinfo {author}
  {\bibfnamefont {B.}~\bibnamefont {Hedquist}},\ and\ \bibinfo {author}
  {\bibfnamefont {A.}~\bibnamefont {Dallman}},\ }\bibfield  {title} {\bibinfo
  {title} {Flow, turbulence, and pollutant dispersion in urban atmospheres},\
  }\href@noop {} {\bibfield  {journal} {\bibinfo  {journal} {Physics of
  Fluids}\ }\textbf {\bibinfo {volume} {22}} (\bibinfo {year}
  {2010})}\BibitemShut {NoStop}%
\bibitem [{\citenamefont {Abuhegazy}\ \emph {et~al.}(2020)\citenamefont
  {Abuhegazy}, \citenamefont {Talaat}, \citenamefont {Anderoglu},\ and\
  \citenamefont {Poroseva}}]{abuhegazy2020numerical}%
  \BibitemOpen
  \bibfield  {author} {\bibinfo {author} {\bibfnamefont {M.}~\bibnamefont
  {Abuhegazy}}, \bibinfo {author} {\bibfnamefont {K.}~\bibnamefont {Talaat}},
  \bibinfo {author} {\bibfnamefont {O.}~\bibnamefont {Anderoglu}},\ and\
  \bibinfo {author} {\bibfnamefont {S.~V.}\ \bibnamefont {Poroseva}},\
  }\bibfield  {title} {\bibinfo {title} {Numerical investigation of aerosol
  transport in a classroom with relevance to covid-19},\ }\href@noop {}
  {\bibfield  {journal} {\bibinfo  {journal} {Physics of Fluids}\ }\textbf
  {\bibinfo {volume} {32}} (\bibinfo {year} {2020})}\BibitemShut {NoStop}%
\bibitem [{\citenamefont {T{\'e}l}\ \emph {et~al.}(2005)\citenamefont
  {T{\'e}l}, \citenamefont {de~Moura}, \citenamefont {Grebogi},\ and\
  \citenamefont {K{\'a}rolyi}}]{tel2005chemical}%
  \BibitemOpen
  \bibfield  {author} {\bibinfo {author} {\bibfnamefont {T.}~\bibnamefont
  {T{\'e}l}}, \bibinfo {author} {\bibfnamefont {A.}~\bibnamefont {de~Moura}},
  \bibinfo {author} {\bibfnamefont {C.}~\bibnamefont {Grebogi}},\ and\ \bibinfo
  {author} {\bibfnamefont {G.}~\bibnamefont {K{\'a}rolyi}},\ }\bibfield
  {title} {\bibinfo {title} {Chemical and biological activity in open flows: A
  dynamical system approach},\ }\href@noop {} {\bibfield  {journal} {\bibinfo
  {journal} {Physics reports}\ }\textbf {\bibinfo {volume} {413}},\ \bibinfo
  {pages} {91} (\bibinfo {year} {2005})}\BibitemShut {NoStop}%
\bibitem [{\citenamefont {Adler}\ \emph {et~al.}(1965)\citenamefont {Adler},
  \citenamefont {Konheim},\ and\ \citenamefont
  {McAndrew}}]{adler1965topological}%
  \BibitemOpen
  \bibfield  {author} {\bibinfo {author} {\bibfnamefont {R.~L.}\ \bibnamefont
  {Adler}}, \bibinfo {author} {\bibfnamefont {A.~G.}\ \bibnamefont {Konheim}},\
  and\ \bibinfo {author} {\bibfnamefont {M.~H.}\ \bibnamefont {McAndrew}},\
  }\bibfield  {title} {\bibinfo {title} {Topological entropy},\ }\href@noop {}
  {\bibfield  {journal} {\bibinfo  {journal} {Transactions of the American
  Mathematical Society}\ }\textbf {\bibinfo {volume} {114}},\ \bibinfo {pages}
  {309} (\bibinfo {year} {1965})}\BibitemShut {NoStop}%
\bibitem [{\citenamefont {Ott}(2002)}]{ott2002chaos}%
  \BibitemOpen
  \bibfield  {author} {\bibinfo {author} {\bibfnamefont {E.}~\bibnamefont
  {Ott}},\ }\bibfield  {title} {\bibinfo {title} {Chaos in dynamical systems},\
  }\href@noop {} {\bibfield  {journal} {\bibinfo  {journal} {Chaos in Dynamical
  Systems-2nd Edition}\ ,\ \bibinfo {pages} {490}} (\bibinfo {year}
  {2002})}\BibitemShut {NoStop}%
\bibitem [{\citenamefont {Newhouse}\ and\ \citenamefont
  {Pignataro}(1993)}]{newhouse1993estimation}%
  \BibitemOpen
  \bibfield  {author} {\bibinfo {author} {\bibfnamefont {S.}~\bibnamefont
  {Newhouse}}\ and\ \bibinfo {author} {\bibfnamefont {T.}~\bibnamefont
  {Pignataro}},\ }\bibfield  {title} {\bibinfo {title} {On the estimation of
  topological entropy},\ }\href@noop {} {\bibfield  {journal} {\bibinfo
  {journal} {Journal of statistical physics}\ }\textbf {\bibinfo {volume}
  {72}},\ \bibinfo {pages} {1331} (\bibinfo {year} {1993})}\BibitemShut
  {NoStop}%
\bibitem [{\citenamefont {Ottino}\ \emph {et~al.}(2004)\citenamefont {Ottino},
  \citenamefont {Wiggins}, \citenamefont {Stremler}, \citenamefont {Haselton},\
  and\ \citenamefont {Aref}}]{doi:10.1098/rsta.2003.1360}%
  \BibitemOpen
  \bibfield  {author} {\bibinfo {author} {\bibfnamefont {J.~M.}\ \bibnamefont
  {Ottino}}, \bibinfo {author} {\bibfnamefont {S.~R.}\ \bibnamefont {Wiggins}},
  \bibinfo {author} {\bibfnamefont {M.~A.}\ \bibnamefont {Stremler}}, \bibinfo
  {author} {\bibfnamefont {F.~R.}\ \bibnamefont {Haselton}},\ and\ \bibinfo
  {author} {\bibfnamefont {H.}~\bibnamefont {Aref}},\ }\bibfield  {title}
  {\bibinfo {title} {Designing for chaos: applications of chaotic advection at
  the microscale},\ }\href {https://doi.org/10.1098/rsta.2003.1360} {\bibfield
  {journal} {\bibinfo  {journal} {Philosophical Transactions of the Royal
  Society of London. Series A: Mathematical, Physical and Engineering
  Sciences}\ }\textbf {\bibinfo {volume} {362}},\ \bibinfo {pages} {1019}
  (\bibinfo {year} {2004})},\ \Eprint
  {https://arxiv.org/abs/https://royalsocietypublishing.org/doi/pdf/10.1098/rsta.2003.1360}
  {https://royalsocietypublishing.org/doi/pdf/10.1098/rsta.2003.1360}
  \BibitemShut {NoStop}%
\bibitem [{\citenamefont {Lee}\ \emph {et~al.}(2011)\citenamefont {Lee},
  \citenamefont {Chang}, \citenamefont {Wang},\ and\ \citenamefont
  {Fu}}]{lee2011microfluidic}%
  \BibitemOpen
  \bibfield  {author} {\bibinfo {author} {\bibfnamefont {C.-Y.}\ \bibnamefont
  {Lee}}, \bibinfo {author} {\bibfnamefont {C.-L.}\ \bibnamefont {Chang}},
  \bibinfo {author} {\bibfnamefont {Y.-N.}\ \bibnamefont {Wang}},\ and\
  \bibinfo {author} {\bibfnamefont {L.-M.}\ \bibnamefont {Fu}},\ }\bibfield
  {title} {\bibinfo {title} {Microfluidic mixing: a review},\ }\href@noop {}
  {\bibfield  {journal} {\bibinfo  {journal} {International journal of
  molecular sciences}\ }\textbf {\bibinfo {volume} {12}},\ \bibinfo {pages}
  {3263} (\bibinfo {year} {2011})}\BibitemShut {NoStop}%
\bibitem [{\citenamefont {Finn}\ and\ \citenamefont
  {Thiffeault}(2011)}]{finn2011topological}%
  \BibitemOpen
  \bibfield  {author} {\bibinfo {author} {\bibfnamefont {M.~D.}\ \bibnamefont
  {Finn}}\ and\ \bibinfo {author} {\bibfnamefont {J.-L.}\ \bibnamefont
  {Thiffeault}},\ }\bibfield  {title} {\bibinfo {title} {Topological
  optimization of rod-stirring devices},\ }\href@noop {} {\bibfield  {journal}
  {\bibinfo  {journal} {SIAM review}\ }\textbf {\bibinfo {volume} {53}},\
  \bibinfo {pages} {723} (\bibinfo {year} {2011})}\BibitemShut {NoStop}%
\bibitem [{\citenamefont {Chu}\ \emph {et~al.}(2018)\citenamefont {Chu},
  \citenamefont {Hu}, \citenamefont {Zhang}, \citenamefont {Gao}, \citenamefont
  {Fang}, \citenamefont {Liu}, \citenamefont {Yan}, \citenamefont {Liu},
  \citenamefont {Sun}, \citenamefont {Peng} \emph
  {et~al.}}]{chu2018topological}%
  \BibitemOpen
  \bibfield  {author} {\bibinfo {author} {\bibfnamefont {Y.}~\bibnamefont
  {Chu}}, \bibinfo {author} {\bibfnamefont {Q.}~\bibnamefont {Hu}}, \bibinfo
  {author} {\bibfnamefont {Y.}~\bibnamefont {Zhang}}, \bibinfo {author}
  {\bibfnamefont {Z.}~\bibnamefont {Gao}}, \bibinfo {author} {\bibfnamefont
  {Z.}~\bibnamefont {Fang}}, \bibinfo {author} {\bibfnamefont {L.}~\bibnamefont
  {Liu}}, \bibinfo {author} {\bibfnamefont {Q.}~\bibnamefont {Yan}}, \bibinfo
  {author} {\bibfnamefont {Y.}~\bibnamefont {Liu}}, \bibinfo {author}
  {\bibfnamefont {S.}~\bibnamefont {Sun}}, \bibinfo {author} {\bibfnamefont
  {G.-D.}\ \bibnamefont {Peng}}, \emph {et~al.},\ }\bibfield  {title} {\bibinfo
  {title} {Topological engineering of photoluminescence properties of
  bismuth-or erbium-doped phosphosilicate glass of arbitrary p2o5 to sio2
  ratio},\ }\href@noop {} {\bibfield  {journal} {\bibinfo  {journal} {Advanced
  Optical Materials}\ }\textbf {\bibinfo {volume} {6}},\ \bibinfo {pages}
  {1800024} (\bibinfo {year} {2018})}\BibitemShut {NoStop}%
\bibitem [{\citenamefont {Thiffeault}(2018)}]{thiffeault2018mathematics}%
  \BibitemOpen
  \bibfield  {author} {\bibinfo {author} {\bibfnamefont {J.-L.}\ \bibnamefont
  {Thiffeault}},\ }\bibfield  {title} {\bibinfo {title} {The mathematics of
  taffy pullers},\ }\href@noop {} {\bibfield  {journal} {\bibinfo  {journal}
  {The Mathematical Intelligencer}\ }\textbf {\bibinfo {volume} {40}},\
  \bibinfo {pages} {26} (\bibinfo {year} {2018})}\BibitemShut {NoStop}%
\bibitem [{\citenamefont {Thiffeault}(2010)}]{10.1063/1.3262494}%
  \BibitemOpen
  \bibfield  {author} {\bibinfo {author} {\bibfnamefont {J.-L.}\ \bibnamefont
  {Thiffeault}},\ }\bibfield  {title} {\bibinfo {title} {Braids of entangled
  particle trajectories},\ }\href {https://doi.org/10.1063/1.3262494}
  {\bibfield  {journal} {\bibinfo  {journal} {Chaos: An Interdisciplinary
  Journal of Nonlinear Science}\ }\textbf {\bibinfo {volume} {20}},\ \bibinfo
  {pages} {017516} (\bibinfo {year} {2010})},\ \Eprint
  {https://arxiv.org/abs/https://pubs.aip.org/aip/cha/article-pdf/doi/10.1063/1.3262494/14604297/017516\_1\_online.pdf}
  {https://pubs.aip.org/aip/cha/article-pdf/doi/10.1063/1.3262494/14604297/017516\_1\_online.pdf}
  \BibitemShut {NoStop}%
\bibitem [{\citenamefont {Candelaresi}\ \emph {et~al.}(2017)\citenamefont
  {Candelaresi}, \citenamefont {Pontin},\ and\ \citenamefont
  {Hornig}}]{candelaresi2017quantifying}%
  \BibitemOpen
  \bibfield  {author} {\bibinfo {author} {\bibfnamefont {S.}~\bibnamefont
  {Candelaresi}}, \bibinfo {author} {\bibfnamefont {D.~I.}\ \bibnamefont
  {Pontin}},\ and\ \bibinfo {author} {\bibfnamefont {G.}~\bibnamefont
  {Hornig}},\ }\bibfield  {title} {\bibinfo {title} {Quantifying the tangling
  of trajectories using the topological entropy},\ }\href@noop {} {\bibfield
  {journal} {\bibinfo  {journal} {Chaos: An Interdisciplinary Journal of
  Nonlinear Science}\ }\textbf {\bibinfo {volume} {27}} (\bibinfo {year}
  {2017})}\BibitemShut {NoStop}%
\bibitem [{\citenamefont {Klapper}\ and\ \citenamefont
  {Young}(1995)}]{klapper1995rigorous}%
  \BibitemOpen
  \bibfield  {author} {\bibinfo {author} {\bibfnamefont {I.}~\bibnamefont
  {Klapper}}\ and\ \bibinfo {author} {\bibfnamefont {L.-S.}\ \bibnamefont
  {Young}},\ }\bibfield  {title} {\bibinfo {title} {Rigorous bounds on the fast
  dynamo growth rate involving topological entropy},\ }\href@noop {} {\bibfield
   {journal} {\bibinfo  {journal} {Communications in mathematical physics}\
  }\textbf {\bibinfo {volume} {173}},\ \bibinfo {pages} {623} (\bibinfo {year}
  {1995})}\BibitemShut {NoStop}%
\bibitem [{\citenamefont {Olascoaga}\ and\ \citenamefont
  {Haller}(2012)}]{olascoaga2012forecasting}%
  \BibitemOpen
  \bibfield  {author} {\bibinfo {author} {\bibfnamefont {M.~J.}\ \bibnamefont
  {Olascoaga}}\ and\ \bibinfo {author} {\bibfnamefont {G.}~\bibnamefont
  {Haller}},\ }\bibfield  {title} {\bibinfo {title} {Forecasting sudden changes
  in environmental pollution patterns},\ }\href@noop {} {\bibfield  {journal}
  {\bibinfo  {journal} {Proceedings of the National Academy of Sciences}\
  }\textbf {\bibinfo {volume} {109}},\ \bibinfo {pages} {4738} (\bibinfo {year}
  {2012})}\BibitemShut {NoStop}%
\bibitem [{\citenamefont {Haszpra}\ and\ \citenamefont
  {T{\'e}l}(2013)}]{Hazpra-S}%
  \BibitemOpen
  \bibfield  {author} {\bibinfo {author} {\bibfnamefont {T.}~\bibnamefont
  {Haszpra}}\ and\ \bibinfo {author} {\bibfnamefont {T.}~\bibnamefont
  {T{\'e}l}},\ }\bibfield  {title} {\bibinfo {title} {Topological entropy: A
  lagrangian measure of the state of the free atmosphere},\ }\href
  {https://doi.org/10.1175/JAS-D-13-069.1} {\bibfield  {journal} {\bibinfo
  {journal} {Journal of the Atmospheric Sciences}\ }\textbf {\bibinfo {volume}
  {70}},\ \bibinfo {pages} {4030 } (\bibinfo {year} {2013})}\BibitemShut
  {NoStop}%
\bibitem [{\citenamefont {Haszpra}\ and\ \citenamefont
  {T{\'e}l}(2011)}]{haszpra2011volcanic}%
  \BibitemOpen
  \bibfield  {author} {\bibinfo {author} {\bibfnamefont {T.}~\bibnamefont
  {Haszpra}}\ and\ \bibinfo {author} {\bibfnamefont {T.}~\bibnamefont
  {T{\'e}l}},\ }\bibfield  {title} {\bibinfo {title} {Volcanic ash in the free
  atmosphere: A dynamical systems approach},\ }in\ \href@noop {} {\emph
  {\bibinfo {booktitle} {Journal of Physics: Conference Series}}},\ Vol.\
  \bibinfo {volume} {333}\ (\bibinfo {organization} {IOP Publishing},\ \bibinfo
  {year} {2011})\ p.\ \bibinfo {pages} {012008}\BibitemShut {NoStop}%
\bibitem [{\citenamefont {Westerweel}\ \emph {et~al.}(2013)\citenamefont
  {Westerweel}, \citenamefont {Elsinga},\ and\ \citenamefont
  {Adrian}}]{westerweel2013particle}%
  \BibitemOpen
  \bibfield  {author} {\bibinfo {author} {\bibfnamefont {J.}~\bibnamefont
  {Westerweel}}, \bibinfo {author} {\bibfnamefont {G.~E.}\ \bibnamefont
  {Elsinga}},\ and\ \bibinfo {author} {\bibfnamefont {R.~J.}\ \bibnamefont
  {Adrian}},\ }\bibfield  {title} {\bibinfo {title} {Particle image velocimetry
  for complex and turbulent flows},\ }\href@noop {} {\bibfield  {journal}
  {\bibinfo  {journal} {Annual Review of Fluid Mechanics}\ }\textbf {\bibinfo
  {volume} {45}},\ \bibinfo {pages} {409} (\bibinfo {year} {2013})}\BibitemShut
  {NoStop}%
\bibitem [{\citenamefont {Küchler}\ \emph {et~al.}(2024)\citenamefont
  {Küchler}, \citenamefont {Ibanez~Landeta}, \citenamefont {Moláček},\ and\
  \citenamefont {Bodenschatz}}]{10.1063/5.0211508}%
  \BibitemOpen
  \bibfield  {author} {\bibinfo {author} {\bibfnamefont {C.}~\bibnamefont
  {Küchler}}, \bibinfo {author} {\bibfnamefont {A.}~\bibnamefont
  {Ibanez~Landeta}}, \bibinfo {author} {\bibfnamefont {J.}~\bibnamefont
  {Moláček}},\ and\ \bibinfo {author} {\bibfnamefont {E.}~\bibnamefont
  {Bodenschatz}},\ }\bibfield  {title} {\bibinfo {title} {Lagrangian particle
  tracking at large reynolds numbers},\ }\href
  {https://doi.org/10.1063/5.0211508} {\bibfield  {journal} {\bibinfo
  {journal} {Review of Scientific Instruments}\ }\textbf {\bibinfo {volume}
  {95}},\ \bibinfo {pages} {105110} (\bibinfo {year} {2024})},\ \Eprint
  {https://arxiv.org/abs/https://pubs.aip.org/aip/rsi/article-pdf/doi/10.1063/5.0211508/20206527/105110\_1\_5.0211508.pdf}
  {https://pubs.aip.org/aip/rsi/article-pdf/doi/10.1063/5.0211508/20206527/105110\_1\_5.0211508.pdf}
  \BibitemShut {NoStop}%
\bibitem [{\citenamefont {Toschi}\ and\ \citenamefont
  {Bodenschatz}(2009)}]{toschi2009lagrangian}%
  \BibitemOpen
  \bibfield  {author} {\bibinfo {author} {\bibfnamefont {F.}~\bibnamefont
  {Toschi}}\ and\ \bibinfo {author} {\bibfnamefont {E.}~\bibnamefont
  {Bodenschatz}},\ }\bibfield  {title} {\bibinfo {title} {Lagrangian properties
  of particles in turbulence},\ }\href@noop {} {\bibfield  {journal} {\bibinfo
  {journal} {Annual review of fluid mechanics}\ }\textbf {\bibinfo {volume}
  {41}},\ \bibinfo {pages} {375} (\bibinfo {year} {2009})}\BibitemShut
  {NoStop}%
\bibitem [{\citenamefont {Francois}\ \emph {et~al.}(2015)\citenamefont
  {Francois}, \citenamefont {Xia}, \citenamefont {Punzmann}, \citenamefont
  {Faber},\ and\ \citenamefont {Shats}}]{braid-2D}%
  \BibitemOpen
  \bibfield  {author} {\bibinfo {author} {\bibfnamefont {N.}~\bibnamefont
  {Francois}}, \bibinfo {author} {\bibfnamefont {H.}~\bibnamefont {Xia}},
  \bibinfo {author} {\bibfnamefont {H.}~\bibnamefont {Punzmann}}, \bibinfo
  {author} {\bibfnamefont {B.}~\bibnamefont {Faber}},\ and\ \bibinfo {author}
  {\bibfnamefont {M.}~\bibnamefont {Shats}},\ }\bibfield  {title} {\bibinfo
  {title} {Braid entropy of two-dimensional turbulence},\ }\href
  {https://doi.org/10.1038/srep18564} {\bibfield  {journal} {\bibinfo
  {journal} {Scientific Reports}\ }\textbf {\bibinfo {volume} {5}},\ \bibinfo
  {pages} {18564} (\bibinfo {year} {2015})}\BibitemShut {NoStop}%
\bibitem [{SM()}]{SM}%
  \BibitemOpen
  \bibfield  {title} {\bibinfo {title} {See supplemental material at},\
  }\href@noop {} {\ }\BibitemShut {NoStop}%
\bibitem [{\citenamefont {Kraichnan}(1967)}]{kraichnan1967inertial}%
  \BibitemOpen
  \bibfield  {author} {\bibinfo {author} {\bibfnamefont {R.~H.}\ \bibnamefont
  {Kraichnan}},\ }\bibfield  {title} {\bibinfo {title} {Inertial ranges in
  two-dimensional turbulence},\ }\href@noop {} {\bibfield  {journal} {\bibinfo
  {journal} {Physics of fluids}\ }\textbf {\bibinfo {volume} {10}},\ \bibinfo
  {pages} {1417} (\bibinfo {year} {1967})}\BibitemShut {NoStop}%
\bibitem [{\citenamefont {Maltrud}\ and\ \citenamefont
  {Vallis}(1991)}]{maltrud1991energy}%
  \BibitemOpen
  \bibfield  {author} {\bibinfo {author} {\bibfnamefont {M.}~\bibnamefont
  {Maltrud}}\ and\ \bibinfo {author} {\bibfnamefont {G.}~\bibnamefont
  {Vallis}},\ }\bibfield  {title} {\bibinfo {title} {Energy spectra and
  coherent structures in forced two-dimensional and beta-plane turbulence},\
  }\href@noop {} {\bibfield  {journal} {\bibinfo  {journal} {Journal of Fluid
  Mechanics}\ }\textbf {\bibinfo {volume} {228}},\ \bibinfo {pages} {321}
  (\bibinfo {year} {1991})}\BibitemShut {NoStop}%
\bibitem [{\citenamefont {Wallace}\ and\ \citenamefont
  {Vukoslav{\v{c}}evi{\'c}}(2010)}]{wallace2010measurement}%
  \BibitemOpen
  \bibfield  {author} {\bibinfo {author} {\bibfnamefont {J.~M.}\ \bibnamefont
  {Wallace}}\ and\ \bibinfo {author} {\bibfnamefont {P.~V.}\ \bibnamefont
  {Vukoslav{\v{c}}evi{\'c}}},\ }\bibfield  {title} {\bibinfo {title}
  {Measurement of the velocity gradient tensor in turbulent flows},\
  }\href@noop {} {\bibfield  {journal} {\bibinfo  {journal} {Annual review of
  fluid mechanics}\ }\textbf {\bibinfo {volume} {42}},\ \bibinfo {pages} {157}
  (\bibinfo {year} {2010})}\BibitemShut {NoStop}%
\bibitem [{\citenamefont {Tsinober}\ \emph {et~al.}(1992)\citenamefont
  {Tsinober}, \citenamefont {Kit},\ and\ \citenamefont
  {Dracos}}]{tsinober1992experimental}%
  \BibitemOpen
  \bibfield  {author} {\bibinfo {author} {\bibfnamefont {A.}~\bibnamefont
  {Tsinober}}, \bibinfo {author} {\bibfnamefont {E.}~\bibnamefont {Kit}},\ and\
  \bibinfo {author} {\bibfnamefont {T.}~\bibnamefont {Dracos}},\ }\bibfield
  {title} {\bibinfo {title} {Experimental investigation of the field of
  velocity gradients in turbulent flows},\ }\href@noop {} {\bibfield  {journal}
  {\bibinfo  {journal} {Journal of Fluid Mechanics}\ }\textbf {\bibinfo
  {volume} {242}},\ \bibinfo {pages} {169} (\bibinfo {year}
  {1992})}\BibitemShut {NoStop}%
\end{thebibliography}%

\end{document}


\title{Supplemental materials: Topological Entropy of Two Dimensional Turbulence}

\author{Amal Manoharan,Sai Subramanian and Ashwin Joy}
\email[]{ashwin@physics.iitm.ac.in}
\affiliation{Department of Physics, Indian Institute of Technology Madras, Chennai - 600036}

\date{\today}
\maketitle

\section{Calculation of the  $\theta$  average}
We need to evaluate the integral in Equation (14), given by
\begin{equation}
  \left<\ln f(\lambda, \theta)  \right>_{\theta} = \frac{2}{\pi}\int_0^{\pi/2} \ln f(\lambda, \theta)  \text{d}\theta 
  \label{first-equ}
\end{equation}
where \(f(\lambda, \theta)\) from Equation (7) is 
\begin{equation*}
f(\lambda, \theta)= \sqrt{e^{2 \lambda \tau} \cos^2\theta + e^{-2 \lambda \tau} \sin^2 \theta}  = e^{\lambda \tau}\cos \theta \sqrt{1+e^{-4\lambda \tau}\tan^2 \theta}.
\end{equation*}

Taking the logarithm of the equation, we get
\begin{equation*}
\ln f(\lambda, \theta)  =  \lambda \tau+ \ln (\cos \theta) + \frac{1}{2} \ln(1+e^{-4\lambda \tau}\tan^2 \theta).
\end{equation*}

Substituting this expression into Equation \ref{first-equ}, we can rewrite it as
\begin{equation}
\frac{2}{\pi}\int_0^{\pi/2} \ln f(\lambda, \theta) \text{d}\theta = \frac{2}{\pi}\int_0^{\pi/2} \lambda \tau \,\text{d} \theta + \frac{2}{\pi}\int_0^{\pi/2} \ln (\cos \theta)\text{d}\theta + \frac{1}{\pi}\int_0^{\pi/2} \ln(1+e^{-4\lambda \tau}\tan^2 \theta) \text{d}\theta.
  \label{expanded}
  \end{equation}

The first two integrals on the right-hand side can be readily evaluated as \(\lambda \tau\) and \(-\ln 2\), respectively. The third integral is evaluated below. Let us denote the third integral as \(I\):
\begin{equation*}
I=\frac{1}{\pi}\int_0^{\pi / 2} \ln \left(1+c \tan^2 \theta\right) d\theta,
\end{equation*}
where \(c=e^{-4\lambda \tau}\). Substituting \(\tan \theta = x\), which gives \(d\theta = \frac{dx}{1+x^2}\), we get
\begin{equation*}
I=\frac{1}{\pi}\int_0^{\infty} \frac{\ln \left(1+c x^2\right)}{1+x^2} dx.
\end{equation*}

This integral belongs to a family of Serret integrals, which can be solved using complex analysis. Consider the following contour integral:
\begin{equation}
\begin{aligned}
J &= \frac{1}{\pi}\oint_C \frac{\ln(z\sqrt{c}+i)}{1+z^2} dz \\
&= 2 i \operatorname{Res}\left[\frac{\ln(z\sqrt{c}+i)}{1+z^2}, z=i\right] \\
&= 2 i \lim_{z \to i} \frac{\ln(z\sqrt{c}+i)}{(z-i)(z+i)} \\
&=  \ln[i(\sqrt{c}+1)] \\
&= \ln e^{i \frac{\pi}{2}} + \ln(\sqrt{c}+1).
\end{aligned}
\label{theta_integral_1}
\end{equation}

\begin{figure}
    \centering
    \includegraphics[scale=0.5]{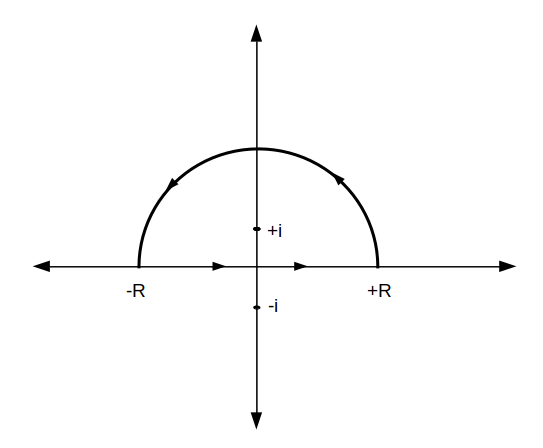}
    \caption{Anti-clockwise contour in the complex plane used to evaluate the Serret integral in Equation \ref{theta_integral_1}, consisting of a semicircular arc of radius \(R\). }
    \label{fig:contour}
\end{figure}

Using the contour integral, we write
\begin{equation}
\begin{aligned}
J &=\frac{1}{\pi} \oint_C \frac{\ln(z\sqrt{c}+i)}{1+z^2} dz \\
&= \frac{1}{\pi} \int_{-R}^R \frac{\ln(x\sqrt{c}+i)}{1+x^2} dx + \frac{1}{\pi} \int_{\theta=0}^\pi \frac{\ln\left(R e^{i\theta}\sqrt{c}+i\right)}{1+R^2 e^{2i\theta}} R i e^{i\theta} d\theta.
\end{aligned}
\end{equation}

As \(R \to \infty\), the second term vanishes:
\begin{equation*}
\left|\frac{\ln\left(R e^{i\theta} \sqrt{c}+i\right) R i e^{i\theta}}{1+R^2 e^{2i\theta}}\right| \to 0.
\end{equation*}

Thus,
\begin{equation}
\begin{aligned}
J &= \frac{1}{\pi}\int_{-\infty}^{\infty} \frac{\ln(x\sqrt{c}+i)}{1+x^2} dx = \frac{1}{\pi}\int_{-\infty}^0 \frac{\ln(x\sqrt{c}+i)}{1+x^2} dx + \frac{1}{\pi}\int_0^{\infty} \frac{\ln(x\sqrt{c}+i)}{1+x^2} dx.
\end{aligned}
\end{equation}

For the integral \(\int_{-\infty}^0\), substitute \(x \to -x\):
\begin{equation}
\begin{aligned}
J &= \frac{1}{\pi} \int_{0}^{\infty} \frac{\ln(-x\sqrt{c}+i)}{1+x^2} d(-x) + \frac{1}{\pi}\int_0^{\infty} \frac{\ln(x\sqrt{c}+i)}{1+x^2} dx \\
&= \frac{1}{\pi}\int_0^{\infty} \frac{\ln(-x\sqrt{c}+i) + \ln(x\sqrt{c}+i)}{1+x^2} dx \\
&= \frac{1}{\pi}\int_0^{\infty} \frac{\ln(-1 - x^2c)}{1+x^2} dx \\
&= \frac{1}{\pi}\int_0^{\infty} \frac{\ln(-1) + \ln(e^{i \pi})}{1+x^2} dx \\
&= \left.i \tan^{-1} x\right|_0^{\infty} + I \\
&= \frac{i\pi}{2} + I.
\end{aligned}
\label{theta_integral_2}
\end{equation}

Comparing Equations \ref{theta_integral_1} and \ref{theta_integral_2}, we find
\begin{equation}
    I=\ln (1+\sqrt{c}) = \ln (1+ e^{-2\lambda \tau}).
    \label{contour-integral}
\end{equation}

Substituting Equation \ref{contour-integral} back into Equation \ref{expanded}, we get the desired result:
\begin{equation}
  \frac{2}{\pi}\int_0^{\pi/2} \ln f(\lambda, \theta)  \text{d}\theta = \ln (\cosh(\lambda \tau)).
  \label{theta-avg}
\end{equation}